\documentclass[modern]{aastex61}

\received{}
\revised{}
\accepted{}
\submitjournal{ApJL}

\shorttitle{Constraints on the Density and Internal Strength of 1I/'Oumuamua}
\shortauthors{McNeill et al.}

\begin{document}

\title{Constraints on the Density and Internal Strength of 1I/'Oumuamua}

\correspondingauthor{Andrew McNeill}
\email{andrew.mcneill@nau.edu}

\author{Andrew McNeill}
\affil{Department of Physics and Astronomy, Northern Arizona University, Flagstaff, AZ 86011, USA}

\author{David E. Trilling}
\affiliation{Department of Physics and Astronomy, Northern Arizona University, Flagstaff, AZ 86011, USA}

\author{Michael Mommert}
\affiliation{Department of Physics and Astronomy, Northern Arizona University, Flagstaff, AZ 86011, USA}

\begin{abstract}

1I/'Oumuamua was discovered by the Panoramic Survey Telescope and Rapid Response System (Pan-STARRS 1) on 19 October 2017. Unlike all previously discovered minor planets this object was determined to have eccentricity $e > 1.0$, suggesting an interstellar origin. Since this discovery and within the limited window of opportunity, several photometric and spectroscopic studies of the object have been made. Using the measured light curve amplitudes and rotation periods we find that, under the assumption of a triaxial ellipsoid, a density range $1500 < \rho < 2800$ kg m$^{-3}$ matches the observations and no significant cohesive strength is required. We also determine that an aspect ratio of $6\pm 1:1$ is most likely after accounting for phase-angle effects and considering the potential effect of surface properties. This elongation is still remarkable but less than some other estimates.\\

\end{abstract}

\keywords{minor planets, asteroids: individual (1I/Ouamuamua) --- techniques: photometric --- methods: statistical}

\section{Introduction} \label{sec:intro}

1I/2017U1 ('Oumuamua) was discovered by Pan-STARRS 1 in October 2017 and subsequently identified as the first ``hyperbolic asteroid'' (\citealt{williams2017}). Its relatively faint H-magnitude along with its extreme orbital eccentricity mean that it was only observable for a relatively short window of opportunity. The current estimated detection rate of hyperbolic asteroids is estimated to be $0.2$ yr$^{-1}$. When the Large Synoptic Survey Telescope begins operations, the detection rate will increase to $1$ yr$^{-1}$ (\citealt{trilling2017}).

Spectroscopic measurements of 1I suggest a surface comparable to that of cometary nuclei and outer solar system objects (\citealt{fitzsimmons2017}; \citealt{masiero2017}). Light curve observations of the object suggest a highly elongated shape with reported values of $\frac{a}{b} = 5.9 \pm 1.0$ from \cite{bolin2017} and $\frac{a}{b} = 10$ from \cite{meech2017}, assuming the object to be a triaxial ellipsoid. In either case this is an unusually elongated object compared to the known shapes and shape distributions of asteroids. Taking for instance the shape distribution for Main Belt Asteroids we find that only $0.003\%$ of objects would be expected to be this elongated \citep{mcneill2016}.

The amplitude of a light curve is strongly affected by the phase angle of the observations. Increased shadowing and scattering effects at high phase angles cause light curve minima to appear fainter and hence the apparent light curve amplitude increases. This can lead to an overestimation of the elongation of an object. 
In \S2 we correct for phase angle effects to derive the most likely
shape of 1I. We then use the assumption of fluid equilibrium
to derive in \S3 the internal cohesive strength of 1I. In \S4 we
discuss our results and some implications of our work.

\section{Phase Angle Effects} \label{sec:phase}

\cite{zappala1990} showed a linear relationship between the apparent amplitude of a light curve $A_{obs}$ and its actual amplitude $A(\alpha=0^{\circ})$ for phase angles $\alpha \leq 40^{\circ}$:

\begin{equation}
A(\alpha=0^{\circ}) = \frac{A_{obs}}{1+s\alpha} .
\label{eqn:phase}
\end{equation}


where $s$ is a taxonomy-dependent slope parameter. Table 1 shows a summary of the phase angles of the reported light curve observations of 1I. All reported light curve observations of 1I fall within this $\alpha$ domain and hence Equation~\ref{eqn:phase} can be used to correct measured light curve amplitudes.

\begin{deluxetable*}{ccccc}[h]
\tablecaption{Summary of the phase angles of reported 1I/'Oumuamua light curve observations considered here
)}
\tablecolumns{5}
\tablenum{1}
\tablewidth{50pt}
\tablehead{
\colhead{Date} &
\colhead{$\alpha$} & \colhead{$A_{obs}$ (mag)} &
\colhead{Telescope} & \colhead{Reference} 
}
\startdata
Oct-25-2017 & $19^{\circ}$ & 2.5 & VLT & \cite{meech2017} \\
Oct-26-2017 & $21^{\circ}$ & 2.5 & GS+VLT & \cite{meech2017} \\
Oct-27-2017 & $22^{\circ}$ & 2.5 & GS+CFHT+UKIRT & \cite{meech2017} \\
Oct-29-2017 & $24^{\circ}$ & $>1.2$ & APO & \cite{bolin2017}\\
Oct-30-2017 & $24^{\circ}$ & $>1.5$ & DCT & \cite{knight2017}\\
\enddata
\label{table:obs}
\end{deluxetable*}

\cite{meech2017} show a light curve amplitude of approximately 2.5 magnitudes: an incredibly large variation corresponding to an elongation of 10:1 if only geometric contributions are considered. From observations made several days later, \cite{bolin2017} report a light curve with amplitude $A_{obs} = 2.05 \pm 0.53$ magnitudes. Again assuming that all variation is purely due to geometric effects, this suggests an elongation between 4:1 and 11:1. At its upper-end this is in agreement with the result of \cite{meech2017}. \cite{bolin2017} use Equation~\ref{eqn:phase} to correct for the $\alpha = 24^{\circ}$ phase angle of their observations and assume a slope parameter $s = 0.015$ mag deg$^{-1}$. This yields a corrected amplitude $A(\alpha=0^{\circ}) = 1.51 \pm 0.39$ mag, corresponding geometrically to an elongation between 3:1 and 6:1.
We apply a similar correction to the \cite{meech2017} $A_{obs}$, which was measured at $\alpha = 22^{\circ}$. This produces an amplitude $A(\alpha=0^{\circ}) = 1.9$ mag, corresponding to an elongation of approximately 6:1. 


Due to 1I's flyby geometry, there have not been enough observations at different orbital geometries to allow light curve inversion to be carried out. 
To understand the uncertainties present in results derived from the data shown in Table~1,
we use the \cite{durech2010} light curve inversion technique code with a fixed spin pole latitude of $\beta = 90^{\circ}$. This model accounts for surface effects as well as geometric effects. In this case we find that the effects of scattering and/or limb darkening can affect the determined elongation of an object by approximately $\frac{a}{b} \pm 1$. This should be considered as an uncertainty in any stated elongation limits derived from a single light curve. Therefore we consider our final result to be $\frac{a}{b}= 6\pm 1$

\section{Constraints on the density and cohesive strength of `Oumuamua} \label{sec:lc}

\subsection{Density}

The critical rotation period of an object $P_{crit}$ is defined at the point where the centripetal force due to rotation is equal to the self-gravity of the object; if the asteroid spins up from this critical rotation period then mass shedding will commence. 

The potential at the surface of an ellipsoid can be represented as Equation~\ref{eqn:potential} where the three $A_{i}$ functions are dimensionless parameters dependent on the axis ratios of the body and are given in Equations~\ref{eqn:Ax}, \ref{eqn:Ay} and \ref{eqn:Az}.

\begin{equation}
\Phi(a, b, c) = -\pi G \rho(A_{0}-A_{x}a^{2}-A_{y}b^{2}-A_{z}c^{2})
\label{eqn:potential}
\end{equation}


\begin{equation} 
A_{x}=\frac{c}{a}\frac{b}{a}\int_{0}^{\infty}\frac{1}{(u+1)^{3/2}(u+\frac{b}{a}^{2})^{1/2}(u+\frac{c}{a}^{2})^{1/2}}du
\label{eqn:Ax}
\end{equation}

\begin{equation} 
A_{y}=\frac{c}{a}\frac{b}{a}\int_{0}^{\infty}\frac{1}{(u+1)^{1/2}(u+\frac{b}{a}^{2})^{3/2}(u+\frac{c}{a}^{2})^{1/2}}du
\label{eqn:Ay}
\end{equation}

\begin{equation} 
A_{z}=\frac{c}{a}\frac{b}{a}\int_{0}^{\infty}\frac{1}{(u+1)^{1/2}(u+\frac{b}{a}^{2})^{1/2}(u+\frac{c}{a}^{2})^{3/2}}du .
\label{eqn:Az}
\end{equation}

By setting the acceleration at the tip of the object, i.e. (x, y, z) = (a, 0, 0) to be equal to the centrifugal acceleration, we can determine the critical angular frequency at which the body will undergo rotational fission. This is given in Equation~\ref{eqn:omegacrit}. This can be rearranged to give the critical rotation period, $P_{crit}$, given in Equation~\ref{eqn:pcrit} where $\rho$ is the density in grams per cubic centimetre.

\begin{equation}
\omega_{crit}=\sqrt{2\pi G\rho A_{x}}
\label{eqn:omegacrit}
\end{equation}

\begin{equation}
P_{crit}=\frac{2.7 \rm{h}}{\sqrt{\rho}}\frac{1}{\sqrt{A_x}} .
\label{eqn:pcrit}
\end{equation}

For spherical objects this becomes $P_{crit}=\frac{3.3 \rm{h}}{\sqrt{\rho}}$. This value is taken by \cite{bolin2017} and scaled according to the $\frac{a}{b}$ axis ratio. This allows them to determine a lower density limit for 1I assuming its rotation period to be equal to the critical rotation period, producing a value $\rho = 1000$ kg m$^{-3}$. This assumption, however, only holds for spherical or near-spherical objects. Instead we determine a more suitable equation for an elongated object assuming $\frac{a}{b}=6$, giving $A_{x}=0.086$. Substituting this into Equation~\ref{eqn:pcrit} we find that for 1I and objects like it a better equation for the lower density limit is given by Equation~\ref{eqn:newlim}.

\begin{equation}
P_{crit}= \frac{9.21 \rm{h}}{\sqrt{\rho}}
\label{eqn:newlim}
\end{equation}

Taking the rotation period determined by \cite{meech2017}, P=$7.34$ h, we determine a lower density limit $\rho_{lim}=1600$ kg m$^{-3}$ This value only represents a lower limit for a cohesionless body, as it assumes that the rotation period is exactly equal to the critical spin rate of the body.




A more sophisticated approach is to assume the object to be a strengthless Jacobi ellipsoid approximating a rubble-pile, a valid assumption for most asteroids with $D\geq 200$ m. (This size is close to the estimated size of 1I so a different
geophysical regime could apply, as discussed below.)
The shape of a Jacobi ellipsoid required to generate a light curve of a given amplitude can be calculated by setting $a=1$ and solving Equation~\ref{eqn:chandraparty} from \cite{chandra1969} for relative axis ratios $b$ and $c$:

\begin{equation} 
a^{2}b^{2}\int_{0}^{\infty}\frac{du}{(a^{2}+u)(b^{2}+u)\Delta}=c^{2}\int_{0}^{\infty}\frac{du}{(c^{2}+u)\Delta}
\label{eqn:chandraparty}
\end{equation}

where $\Delta$ is defined by

\begin{equation} 
\Delta^{2}=(a^{2}+u)(b^{2}+u)(c^{2}+u) .
\label{eqn:Delta}
\end{equation}

When $abc$ is known and the angular rotation frequency of the asteroid, $\omega$, 
is also known then the density may be estimated (also from \cite{chandra1969}):

\begin{equation} 
\frac{\omega^{2}}{\pi G\rho}=2abc\int_{0}^{\infty}\frac{u du}{(a^{2}+u)(b^{2}+u)\Delta} .
\label{eqn:density2}
\end{equation}

Here $G$ is the gravitational constant and $\rho$ is the density of the body in kg m$^{-3}$; it is assumed that the density of the object is constant throughout and that there is no internal strength.
An object with an amplitude $A_{obs} = 1.9$ mag produces a best fit with $a:b:c$ axis ratios $1:0.17:0.16$ and a density of $1800< \rho < 2200$ kg m$^{-3}$, consistent with the $\rho_{lim}$ determined previously.

\subsection{Internal strength}



\cite{jeans1919} states that strengthless ellipsoids where $\frac{b}{a} < 0.44$ are potentially unstable;
our solution has $\frac{b}{a}$ of around~0.17. They assume the ellipsoid to be an incompressible fluid. Highly elongated fluid objects will settle into less elongated equilibrium shapes due to fluid instability. There is a limit to how well a fluid ellipsoid approximates a rubble pile asteroid as these will have some internal friction, which will affect the equilibrium end-states of the object. \cite{holsapple2001} states that for known asteroids assuming "a modest angle of friction" elongated shapes can be maintained and these fluid instabilities are negligible.
Therefore we attempt to calculate the cohesive strength required using a simplified Drucker-Prager model (\citealt{holsapple2004}).
The Drucker-Prager failure criterion is a model of the three-dimensional stresses within a geological material at its critical rotation state. The shear stresses on a body, $\sigma$, in three orthogonal $xyz$ axes are dependent on the shape, density, and rotational properties of the body (\citealt{holsapple2007}): 

\begin{equation} 
\sigma_{x}=(\rho\omega^{2}-2\pi\rho^{2}GA_{x})\frac{a^{2}}{5}
\label{eqn:sigmax}
\end{equation}

\begin{equation} 
\sigma_{y}=(\rho\omega^{2}-2\pi\rho^{2}GA_{y})\frac{b^{2}}{5}
\label{eqn:sigmay}
\end{equation}

\begin{equation} 
\sigma_{z}=(-2\pi\rho^{2}GA_{z})\frac{c^{2}}{5} .
\label{eqn:sigmaz}
\end{equation}

The Drucker-Prager failure criterion is the point at which the object will undergo rotational fission and is given by:

\begin{equation} 
\frac{1}{6}[(\sigma_{x}-\sigma_{y})^{2}+(\sigma_{y}-\sigma_{z})^{2}+(\sigma_{z}-\sigma_{x})^{2}] \leq [k-s(\sigma_{x}+\sigma_{y}+\sigma_{z})]^{2}
\label{eqn:drucker}
\end{equation}

\noindent where $k$ is the internal cohesive strength of the body and $s$ is a slope parameter dependent on the assumed angle of friction, $\phi$:

\begin{equation} 
s=\frac{2\rm{sin}\phi}{\sqrt{3}(3-\rm{sin}\phi)} .
\label{eqn:slope}
\end{equation}

All asteroids modeled by \cite{holsapple2004} have an angle of friction, $\phi_{F}<40^{\circ}$ with most having $\phi_{F}<15^{\circ}$. We assume the angle of friction for 1I in this case to be $\phi = 15^{\circ}$.

Using a simple model based on this failure criterion we determine the required cohesive strength as a function of density for 1I using input parameters determined from the observations. We use a Monte Carlo numerical simulation to determine the required strengths for a range of synthetic objects generated using the estimates of size, shape and rotation period of 1I and their associated uncertainties. This is repeated for a wide range of possible densities. For an object of 1I's estimated size and elongation and assuming sensible density estimates we find that a cohesive strength of only a few Pascals is required --- essentially no significant cohesive strength. This is in agreement with the result presented by \cite{bolin2017}.

Using Equation~\ref{eqn:drucker} it is possible to set constraints on the density of an object assuming zero cohesive strength, i.e. $k=0$. Assuming an angle of friction $\phi_{F}=15^{\circ}$ and P=7.34 h we determined that for an object with 1I's estimated shape a density range $1500 < \rho < 2800$ kg m$^{-3}$ is found. This density range is consistent with the assumption that 1I is a rubble pile.

\section{Discussion and caveats}

The elongations determined for 1I are based upon the assumption that the light curve amplitude is entirely due to shape (not variations in surface reflectivity; see below) and that the object is being viewed equatorially. From the existing data, no period solution has been entirely agreed upon. \cite{bannister2017} determine a period solution of $8.1$ h, while \cite{meech2017} from a full light curve determine a period of $7.34$ h. It has been proposed by \cite{fraser2017} that the seemingly variable rotation rate is due to non-principal axis rotation. This would explain why it has not been possible for a single period to fit all of the light curves. If this is  the case it is difficult to assess the validity of the assumption that the object was observed equator-on at its maximum amplitude. This means that the elongation is a lower limit and hence the density and strength estimates may be underestimated.


The V-shaped light curve minima observed from 1I could be an indication of binarity (\citealt{knight2017}, \citealt{thirouin2017}). From the observations obtained for the object to date, however, it is not possible to determine if this is the case. In the case of a binary system, the required density is effectively the same as that required for a Jacobi ellipsoid. The high amplitude of this asteroid's light curve can be explained either with a single, highly elongated body of $\frac{a}{b} \geq 6$ or a binary system. Asteroids with light curve amplitudes similar to that displayed by 1I (e.g., (1620) Geographos) are generally explained using highly elongated single bodies and we consider this to be the simpler and more likely option.

If the object is monolithic in nature then the fluid approximations used here will be invalid. This requires a higher density due to lack of porosity, and the lower density limit estimate is still valid.

It is also possible to produce a large light curve amplitude if there is a significant variation in the albedo of the body across its surface. However, there are no minor planets known to have such a large variation. The only object in the Solar System with such a variation is the Saturnian moon Iapetus, which is highly variegated due to the fact that it sweeps up dust preferentially on one
hemisphere.

The estimated density of $2000$ kg m$^{-3}$ is consistent with the previous average density estimates for asteroids (\citealt{carry2012}). This density is greater than that expected for cometary nuclei. The density value obtained is also less than that of most meteorite samples, suggesting that 1I must have some degree of porosity which supports a rubble pile assumption. The small required cohesive strength, of order several Pascals, suggests that if 1I is a rubble pile that it is effectively strengthless. Rubble pile asteroids have been determined to have possible strengths from zero to several hundred Pascals, so in this respect 1I is not unusual (\citealt{polishook2016}). It is worth noting that the assumption that this object is a rubble pile is valid assuming $D > 150$ m (\citealt{pravec2002}). For smaller diameter objects a monolithic structure is more likely.

\section{Conclusions} \label{sec:conclusions}

Using the reported light curve amplitudes and rotation periods of the hyperbolic asteroid 1I/2017U1 ('Oumuamua) 
and accounting for phase-amplitude effects we determine a lower limit on the elongation of 1I of~6:1.
Assuming a triaxial ellipsoid we constrain the plausible density range of this object to be $1500 < \rho < 2800$ kg m$^{-3}$; no significant cohesive strength is required at this density. It is possible to obtain a valid binary density solution for the system but there is currently no evidence to favor this over the single-body explanation. These values are based on the assumption that the object was observed equatorially. If the object is tumbling, as proposed by \cite{fraser2017}, this is a more complicated scenario. As such it should be emphasised that the elongation determined for the object is a lower limit, which may then lead to an underestimation of the density and cohesive strength of the object.

\acknowledgments
This work is supported in part by
NSF award 1229776
and NASA award
NNX12AG07G. We thank John Dubinski and an anonymous referee for their insight which has improved the overall content of the paper.

\vspace{5mm}
\facilities{}

\software{Light curve inversion source code (\citealt{durech2010})}

\end{document}